\documentclass[journal]{IEEEtran}
\newcommand{\RNum}[1]{\uppercase\expandafter{\romannumeral #1\relax}}
\usepackage{mathrsfs}
\usepackage{amssymb}
\usepackage[mathcal]{euscript}
\usepackage{diagbox}
\usepackage{cite}
\usepackage[T1]{fontenc}
\usepackage{graphicx}
\usepackage{CJKutf8}
\usepackage{makecell}
\usepackage{psfrag}
\usepackage{url}
\usepackage{stfloats}
\usepackage{amsmath}
\usepackage{array}
\usepackage{float}
\usepackage{multirow} 
\usepackage{hyperref}
\usepackage{amsthm}
\usepackage{color}
\usepackage{multicol}
\usepackage{stfloats}
\usepackage{enumerate}
\usepackage{subfigure}
\usepackage{booktabs}  

\allowdisplaybreaks[4]
\usepackage[ruled]{algorithm2e}
\usepackage{enumitem}
\usepackage{algorithmic} 
\usepackage{xurl}

\ifCLASSINFOpdf
\else
\fi
\begin{document}
\begin{CJK}{UTF8}{gbsn}
%
\title{Toward Agentic AI: Generative Information Retrieval Inspired Intelligent Communications and Networking}



%
	\author{Ruichen Zhang, Shunpu Tang, Yinqiu Liu, Dusit Niyato,~\IEEEmembership{Fellow,~IEEE},  Zehui Xiong, \\ Sumei Sun,~\IEEEmembership{Fellow,~IEEE}, 
Shiwen Mao,~\IEEEmembership{Fellow,~IEEE}, and Zhu Han,~\IEEEmembership{Fellow,~IEEE}

\thanks{R. Zhang, S. Tang, Y. Liu, and D. Niyato are with the College of Computing and Data Science, Nanyang Technological University, Singapore (e-mail: ruichen.zhang@ntu.edu.sg, n2409411h@e.ntu.edu.sg, yinqiu001@e.ntu.edu.sg, dniyato@ntu.edu.sg).}

\thanks{Z. Xiong is with the Computer Science and Design Pillar, University of
Technology and Design, Singapore (e-mail: zehui\_xiong@sutd.edu.sg). }

\thanks{S. Sun is with the Institute for Infocomm Research, Agency for Science, Technology and Research, Singapore (e-mail: sunsm@i2r.a-star.edu.sg). }

\thanks{S. Mao is with the Department of Electrical and Computer Engineering,
Auburn University, Auburn, AL 36849, USA (e-mail: smao@ieee.org).}

\thanks{Z. Han is with the University of Houston, Houston TX 77004, USA, and also with the Department of Computer Science and Engineering, Kyung Hee University, Seoul 446701, South Korea (e-mail: hanzhu22@gmail.com).}
}

\maketitle

\begin{abstract}
The increasing complexity and scale of modern telecommunications networks demand intelligent automation to enhance efficiency, adaptability, and resilience. {Agentic AI has emerged as a key paradigm for intelligent communications and
networking, enabling AI-driven agents to perceive, reason, decide, and act within dynamic networking environments. However, effective decision-making in telecom applications, such as network planning, management, and resource allocation, requires integrating retrieval mechanisms that support multi-hop reasoning, historical cross-referencing, and compliance with evolving 3GPP standards.} This article presents a forward-looking perspective on generative information retrieval-inspired intelligent communications and networking, emphasizing the role of knowledge acquisition, processing, and retrieval in agentic AI for telecom systems. We first provide a comprehensive review of generative information retrieval strategies, including traditional retrieval, hybrid retrieval, semantic retrieval, knowledge-based retrieval, and agentic contextual retrieval. We then analyze their advantages, limitations, and suitability for various networking scenarios. Next, we present a survey about their applications in communications and networking. Additionally, we introduce an agentic contextual retrieval framework to enhance telecom-specific planning by integrating multi-source retrieval, structured reasoning, and self-reflective validation. Experimental results demonstrate that our framework significantly improves answer accuracy, explanation consistency, and retrieval efficiency compared to traditional and semantic retrieval methods. Finally, we outline future research directions.
\end{abstract}


\section{Introduction}

According to a Cisco report, the number of connected devices is expected to surpass 125 billion by 2030\footnote{\url{https://blogs.cisco.com/industrial-iot/iot-is-creating-massive-growth-opportunities}}, requiring networking systems to process massive amounts of data while maintaining seamless interactions across diverse, heterogeneous infrastructures. To support this evolution, modern networks must incorporate intelligent decision-making mechanisms that enable autonomous control, adaptive resource management, and real-time optimization \cite{10599120}. Agentic AI has emerged as a promising paradigm for autonomous network intelligence, addressing the limitations of traditional rule-based and static AI architectures. Introduced by OpenAI\footnote{\url{https://openai.com/}}, DeepSeek\footnote{\url{https://www.deepseek.com/}}, and other research institutions, agentic AI refers to autonomous agents that can perceive, reason, act, and continuously learn from their environments, allowing them to dynamically optimize network configurations, manage resources, and mitigate failures in large-scale systems \cite{sivakumar2024agentic}. Unlike conventional AI, which operates on fixed rules or pre-trained models, agentic AI leverages large language models (LLMs), generative AI-based decision-making, and multi-embodied AI agent collaboration to facilitate self-organizing, highly adaptive network architectures \cite{10679152}. For example,  in \cite{dev2025advanced}, the authors explored intent-based networking with agentic AI, where autonomous agents dynamically updated network management policies based on user-defined intents, achieving a 32\% improvement in QoS requirements and a 40\% reduction in manual intervention for network reconfiguration. Despite its potential, agentic AI faces critical limitations, particularly in handling large-scale network data, maintaining long-term memory, and retrieving historical insights for enhanced decision-making.  Specifically, LLM-based agents often lack efficient information retrieval methods, resulting in hallucinations, context drift, and response inconsistency, which undermine their reliability in real-world networking applications.

To mitigate these limitations, generative information retrieval has been proposed as a fundamental enhancement for agentic AI-driven network intelligence \cite{singh2025agentic}. Unlike traditional retrieval techniques, which rely on static keyword searches and limited contextual matching, generative information retrieval dynamically retrieves, synthesizes, and integrates multi-source knowledge, enabling memory-augmented, context-aware reasoning. For instance, in real-world networking applications, retrieval-augmented AI systems can access historical network logs, regulatory standards, and prior optimization strategies, allowing them to infer multi-hop dependencies across diverse network data sources \cite{anupam2025llm}. This approach significantly enhances decision accuracy, adaptability, and long-term contextual understanding. An example of generative information retrieval in practice is Meta AI’s LlamaIndex\footnote{\url{https://gpt-index.readthedocs.io/en/latest/}}, which enables structured document retrieval for LLM-based applications. It allows AI agents to process and integrate domain-specific knowledge in real-time.

Building on these foundations, this article provides a forward-looking perspective on agentic contextual retrieval and its role in enhancing information retrieval and decision-making within 3GPP-driven autonomous networking environments. Unlike conventional retrieval-augmented AI frameworks, the proposed approach integrates multi-source retrieval, structured reasoning, and self-reflective validation, thereby ensuring improved retrieval accuracy, contextual coherence, and decision consistency. \textit{To the best of our knowledge, this is the first work to explore the potential of agentic contextual retrieval for 3GPP-based telecommunications troubleshooting and real-time standard-compliant decision-making.} The key contributions of this work are summarized as follows.

Firstly, we summarize different retrieval strategies, including traditional retrieval, hybrid retrieval, semantic retrieval, knowledge-based retrieval, and demonstrate the most advanced agentic contextual retrieval. We analyze their applications in networking environments, identifying key challenges and the role of retrieval in enhancing network intelligence. Secondly, we provide a comprehensive review of retrieval-based methodologies in networking and communications, categorizing existing works based on their scenarios, proposed techniques, and publication timelines. This analysis highlights research trends and the evolving role of retrieval in intelligent communications and networking. Finally, we introduce an LLM-based framework that integrates agentic contextual retrieval to improve telecom-specific planning and decision-making. This framework incorporates multi-source knowledge retrieval, reasoning-based decision augmentation, and contextual adaptation, leading to substantial improvements in network optimization, fault diagnosis, and adaptive policy enforcement.

\section{Different Retrieval Methods for Networking}

\begin{figure*}[!t]
\centering
\includegraphics[width=0.99\textwidth]{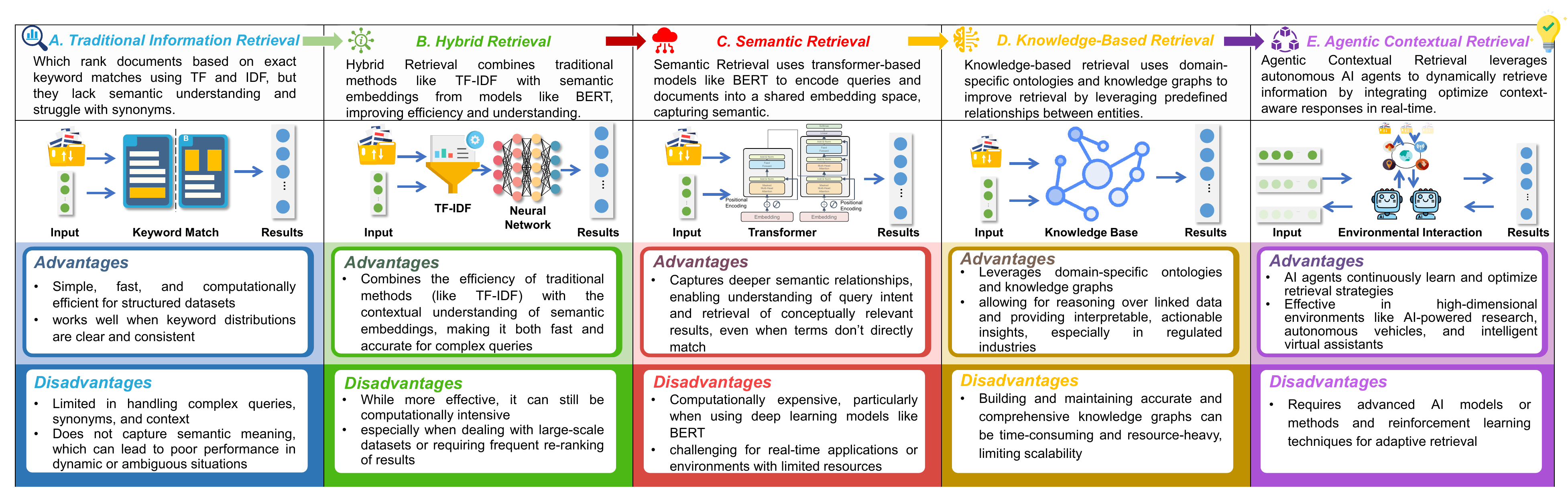}
\caption{Overview of key retrieval strategies in networking. The figure highlights the methodologies, key components, and applications of different approaches, including traditional retrieval, hybrid retrieval, semantic retrieval, knowledge-based retrieval, and agentic contextual retrieval.}
\label{fig_PSNC_1}
\end{figure*}

In intelligent networking, retrieval systems help process vast amounts of unstructured data, optimize spectrum usage, and support AI-based network controllers \cite{singh2025agentic}. In edge intelligence, retrieval techniques facilitate distributed learning, enhance federated AI models, and provide real-time recommendations with minimal latency. As shown in Fig~\ref{fig_PSNC_1}, retrieval methods have evolved from traditional keyword-based approaches to hybrid and context-aware techniques, each addressing specific challenges in networking environments. 

\begin{table*}[!t]
\centering
\caption{Comparison of Key Retrieval Strategies.} 
\includegraphics[width=0.95\textwidth]{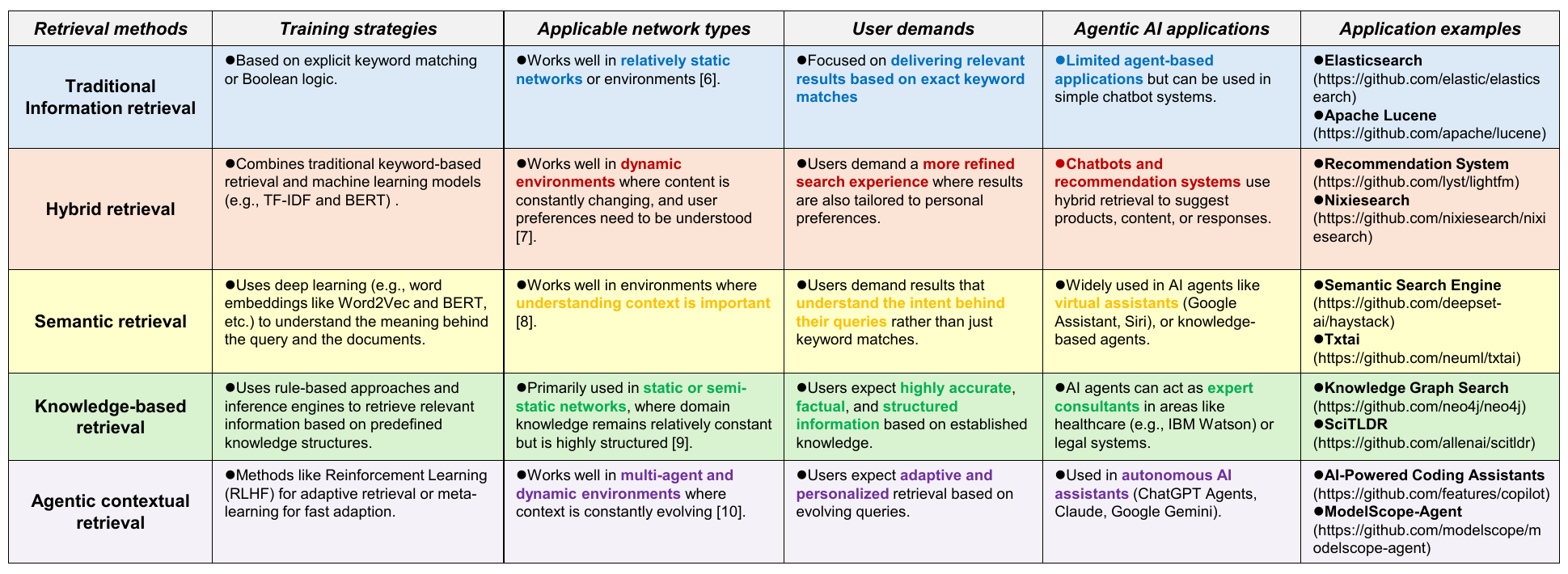}
\label{table_generative AI}
\end{table*}

\subsection{Traditional Information Retrieval}
Traditional information retrieval is based on matching query terms with exact keywords in the dataset, often using simple yet effective algorithms such as Boolean matching or vector space models. These methods calculate document relevance by scoring terms according to their frequency within a document (i.e., term frequency, TF) and across the entire dataset (i.e., inverse document frequency, IDF). The resulting relevance scores rank documents based on their alignment with the query. This approach works well in structured datasets with clear and consistent keyword distributions, such as early library catalog systems or archival searches. However, it does not account for the semantic meaning of terms or the broader context in which the query occurs. To address such issues, for example, Salton et al. \cite{10531073} proposed a foundational vector space model where documents and queries are represented as vectors in a multi-dimensional space. The similarity between these vectors is computed using cosine similarity, allowing for efficient ranking of documents based on query relevance. Experimental results demonstrated that the vector space model improved retrieval precision by 15\% compared to basic Boolean retrieval methods. However, when applied to dynamic datasets such as network resource management logs, its reliance on exact matches caused about a 20\% drop in recall for queries involving synonyms or contextually related terms. These limitations highlight the need for more adaptive retrieval methods in real-time scenarios.

\subsection{Hybrid Retrieval}
Hybrid retrieval combines traditional retrieval methods, such as TF-IDF scoring, with semantic embeddings generated by pre-trained deep learning models such as BERT or GPT. This hybrid approach addresses the limitations of traditional methods by incorporating contextual understanding while maintaining computational efficiency. In hybrid retrieval, the process typically contains two stages: \textit{a coarse filtering stage}, which uses lightweight traditional methods to identify a subset of candidate documents, followed by \textit{a re-ranking stage} where semantic embeddings are applied to refine results. This two-stage approach ensures that hybrid retrieval is both efficient and accurate, making it particularly suitable for environments where computational resources are limited but semantic depth is required. In networking applications, hybrid retrieval can be particularly useful for AI-driven network monitoring and anomaly detection, where efficient pre-filtering combined with deep learning enables fast yet context-aware decision-making. For example, Zeng et al. \cite{zeng2024federated} proposed a federated hybrid retrieval framework designed to integrate traditional TF-IDF filtering with semantic re-ranking using BERT embeddings. Their system processed candidate documents in two stages: first, TF-IDF was used to rapidly filter out irrelevant data at mobile edge nodes, significantly reducing the search space; second, the filtered candidates were semantically ranked using embeddings. Experimental results showed that this approach improved retrieval precision by 25\% and reduced computational latency by 20\% compared to other classical retrieval systems.

\subsection{Semantic Retrieval}
Semantic retrieval uses deep neural networks, particularly transformer-based architectures such as BERT, to encode queries and documents into a shared embedding space. This embedding space captures the semantic relationships between terms, enabling the retrieval system to understand the intent behind the query rather than relying solely on exact keyword matches. Semantic retrieval excels in handling complex queries that involve ambiguous or domain-specific language, such as medical diagnostics and network troubleshooting.  For example, Tang et al. \cite{tang2024retrieval} proposed a semantic retrieval framework leveraging BERT-based embeddings to optimize resource allocation in wireless networks. By encoding queries and documents into a shared semantic space, the system retrieved contextually related documents even for complex queries such as ``dynamic spectrum sharing in 5G". Their experiments demonstrated a 32\% increase in recall compared to hybrid retrieval methods and an 18\% improvement in precision. 

\subsection{Knowledge-Based Retrieval}
Knowledge-based retrieval integrates domain-specific ontologies and structured knowledge graphs to enhance retrieval performance. These systems excel in reasoning tasks by explicitly leveraging predefined relationships between entities, providing interpretable results that are often critical in regulated domains such as healthcare, finance, and telecommunications. In knowledge-based retrieval, it is performed by querying the knowledge graph to extract entities and their relationships that match the query context. This method allows for reasoning over linked data, enabling the retrieval of not just relevant documents but also actionable insights based on the relationships in the dataset. For example, Xiong et al. \cite{xiong2024graph} proposed a knowledge graph-based retrieval system for wireless spectrum management. Their framework utilized a graph structure where nodes represented entities such as ``spectrum bands,'' ``user demands,'' and ``interference levels,'' while edges captured relationships such as ``interferes with'' or ``assigned to.''  The key advantage of this approach lies in its ability to provide structured, explainable decisions based on predefined rules. The system achieved a 25\% improvement in spectrum allocation efficiency and a 30\% reduction in interference conflicts compared to heuristic-based methods.

\subsection{Agentic Contextual Retrieval}
Agentic contextual retrieval leverages intelligent agent-based control mechanisms to dynamically adjust retrieval strategies based on task-specific requirements, multimodal data integration, and real-time environmental changes. Unlike traditional or semantic retrieval methods, which rely on static queries and predefined indexing, this approach enables adaptive, goal-driven information extraction that continuously refines itself based on evolving conditions. By incorporating real-time system states, historical patterns, and structured knowledge representations, agentic contextual retrieval ensures high adaptability and context-aware decision-making, making it particularly suited for applications in network optimization, autonomous systems, and intelligent fault diagnostics. A key advantage of agentic contextual retrieval is its ability to enable autonomous decision-making agents that actively monitor, retrieve, and reason over multiple data sources to enhance performance in complex, dynamic environments. For example, Kagaya et al. \cite{kagaya2024rap} proposed a retrieval framework for autonomous driving, where an agent-driven control mechanism integrated LiDAR, GPS, real-time traffic updates, and weather conditions to dynamically adjust navigation strategies. By enabling real-time, intelligent retrieval and control, their system reduced recalibration time by 40\% and improved navigation accuracy by 28\%.

\begin{figure*}[!t]
\centering
\includegraphics[width=0.95\textwidth]{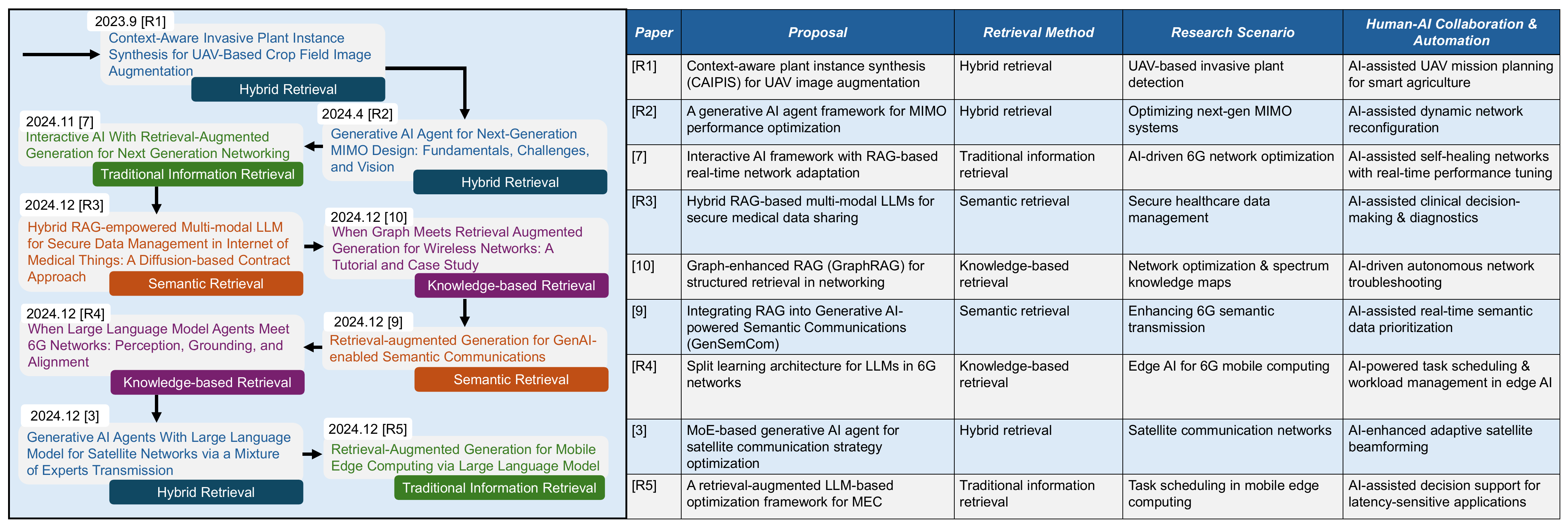}
\caption{A summary of recent retrieval methods in communications and networking, which provides an overview of various proposals, research scenarios, and levels of human-AI interaction.}
\label{fig_PSNC_2}
\end{figure*}

\subsection{Retrieval Comparison and Lessons Learned}
Retrieval methods vary significantly in their methodologies, applications, and suitability for different networking scenarios. {Specifically, traditional retrieval, which relies on explicit keyword matching, is well-suited for static local network management, where queries are simple, computational resources are limited, and speed is prioritized. Hybrid retrieval combines keyword-based search with machine learning models, making it effective for dynamic network environments, such as adaptive caching or content distribution, where user preferences evolve over time. Semantic retrieval, powered by deep learning models, enhances intent-driven network diagnostics by capturing query context, making it particularly useful for automated fault detection and troubleshooting in telecom networks. Knowledge-based retrieval, leveraging structured inference models, supports rule-based network security and access control, where highly accurate, structured decision-making is critical. Finally, agentic contextual retrieval offers adaptive and real-time decision support in multi-agent network control systems, where dynamic environmental factors, such as interference levels or traffic congestion, require continuous learning and adjustment\cite{kagaya2024rap}.} Table~\ref{table_generative AI} summarizes these strategies, highlighting their core features, training methods, and example applications.

Moreover, we conduct a review of recent retrieval-based approaches in communications and networking from 2023 to late 2024, as summarized in Fig.~\ref{fig_PSNC_2}. Our analysis categorizes retrieval strategies into traditional, hybrid, semantic, knowledge-based, and agentic contextual retrieval, highlighting their applications across various domains, including wireless communications, network optimization, and intelligent decision-making. While retrieval-augmented methods have been increasingly integrated into AI-driven network resource management and semantic communication, we observe that agentic contextual retrieval remains largely unexplored for telecommunications-specific applications. Moreover, while \cite{kagaya2024rap} demonstrates agentic contextual retrieval for autonomous driving control, there is currently no direct implementation tailored for communication networks and telecom infrastructure. To fill this gap, the next section introduces our proposed framework, which leverages agentic contextual retrieval to enhance intelligent decision-making, troubleshooting, and autonomous adaptation in telecommunications and networking systems.
\begin{figure*}[!t]
\centering
\includegraphics[width=.95\textwidth]{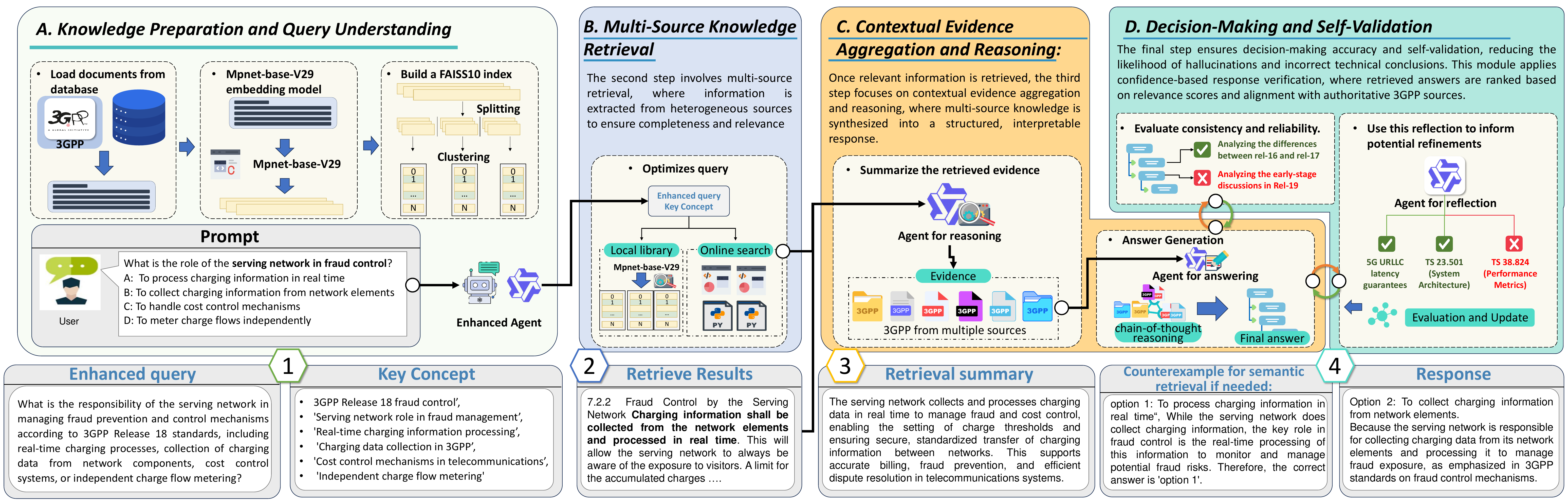}
\caption{Illustration of the agentic contextual retrieval \textcolor{black}{enhanced intelligent base station} for troubleshooting and decision-making. The framework follows a structured four-step workflow: (A) Query understanding and reformulation ensure alignment with 3GPP terminology using LLM-based query expansion. (B) Multi-source knowledge retrieval extracts relevant information from both structured (e.g., 3GPP standards) and unstructured (e.g., online sources) datasets. (C) Contextual evidence aggregation and reasoning synthesize retrieved knowledge into structured responses using chain-of-thought reasoning. (D) Decision-making and self-validation enhance accuracy through confidence-based verification and iterative refinement, reducing hallucinations and improving response consistency.}
\label{fig_PSNC_4}
\end{figure*}

\section{Case Study: Agentic Contextual Retrieval for Networking }

\subsection{Motivation}

In next-generation communications and networking, efficient resource allocation, adaptive service provisioning, and intelligent decision-making are crucial for optimizing user experience and network efficiency. {Modern communication systems are shifting towards intent-driven networking, where mobile users express high-level requirements in natural language, and the network autonomously interprets and executes these requests. However, this paradigm introduces significant challenges in bridging the gap between user intents, structured communication standards, and real-time network configurations. A key challenge lies in mapping natural language intent descriptions to actionable network configurations, requiring an understanding of both human semantics and telecommunications-specific knowledge. Traditional rule-based methods or static intent templates are insufficient in handling diverse user demands and evolving network conditions \cite{dev2025advanced}. LLMs offer a promising solution due to their strong natural language understanding (NLU) and reasoning capabilities. However, LLMs lack domain-specific knowledge in telecommunications, such as 3GPP standards, intent translation templates, and network control logic.} Consequently, their direct application to network automation remains limited by knowledge incompleteness, retrieval inefficiency, and contextual inconsistency.

To address these challenges, we propose a retrieval-enhanced intelligent base station architecture, where the network dynamically retrieves, synthesizes, and applies knowledge from 3GPP standards, network logs, and external telecom repositories to enhance decision-making. \textcolor{black}{Specifically, the system employs a hybrid retrieval framework to convert user-generated intents into structured network actions, using a template-based approach that aligns with communication paradigms outlined in 3GPP \cite{maatouk2023teleqna}. In this framework, user requests (e.g., ``I need ultra-low latency for cloud gaming'') are processed by the network's AI module, which retrieves relevant telecom policies and configurations before generating a customized communication plan. Despite the advantages of retrieval-augmented LLMs, conventional retrieval-augmented generation (RAG) techniques face critical limitations in telecom-specific applications, including: (i) Contextual Ambiguity: Simple keyword-based retrieval struggles to retrieve relevant 3GPP policies and network parameters, as user intents often involve multiple layers of contextual interpretation.
(ii) Data Sparsity: Telecommunications standards and policy documents are highly structured, yet spread across multiple releases and fragmented into different standardization documents.
(iii) Retrieval Inefficiency: Traditional retrieval approaches lack multi-hop reasoning, failing to link user intents with both historical network behavior and real-time conditions.}

\textcolor{black}{To overcome these limitations, we introduce an agentic contextual retrieval framework, which integrates multi-source knowledge retrieval, structured reasoning, and self-reflective validation to enhance intent-driven networking. Our framework enables intelligent base stations to map user intents to network configurations in real-time, leveraging LLM-powered decision-making while ensuring alignment with 3GPP compliance, traffic optimization strategies, and real-world deployment policies.}

\subsection{Agentic Contextual Retrieval Framework}
As shown in Fig.~\ref{fig_PSNC_4}, the deployment of the agentic contextual retrieval framework follows a structured four-step workflow, designed to enhance the retrieval, reasoning, and validation of knowledge specific to 3GPP standards and telecommunications networks.

\subsubsection{\textbf{Knowledge Preparation and Query Understanding}} The system first loads 3GPP standards and network documentation from a database, segments them into context-aware knowledge chunks, and vectorizes them using sentence-transformer embeddings. To enable efficient semantic retrieval, the vectorized knowledge chunks are indexed using a vector database, allowing for efficient similarity searches. After that,  once a query is received, the system analyzes user intent and performs query reformulation, ensuring that the query aligns with 3GPP-defined communication paradigms and technical configurations. In practice, telecommunications queries often contain ambiguous terms, incomplete phrasing, or require historical cross-referencing across multiple 3GPP releases. \textcolor{black}{Therefore, it is necessary to fully understand the user intent and the key concepts in this context to improve retrieval accuracy. Specifically, we can use LLMs to realize that and ensure longitudinal consistency when retrieving regulatory and technical specifications \cite{li2025search}.} In our experimental setup, \textcolor{black}{user intent queries, such as customized communication service requests (e.g., ``I need ultra-reliable low-latency communication for industrial automation''), are first parsed and the key concepts such as ``ultra-reliable low-latency'', ``role of URLLC in industrial automation'' are extracted.}  

\begin{figure*}[!t]
\centering
\includegraphics[width=.85\textwidth]{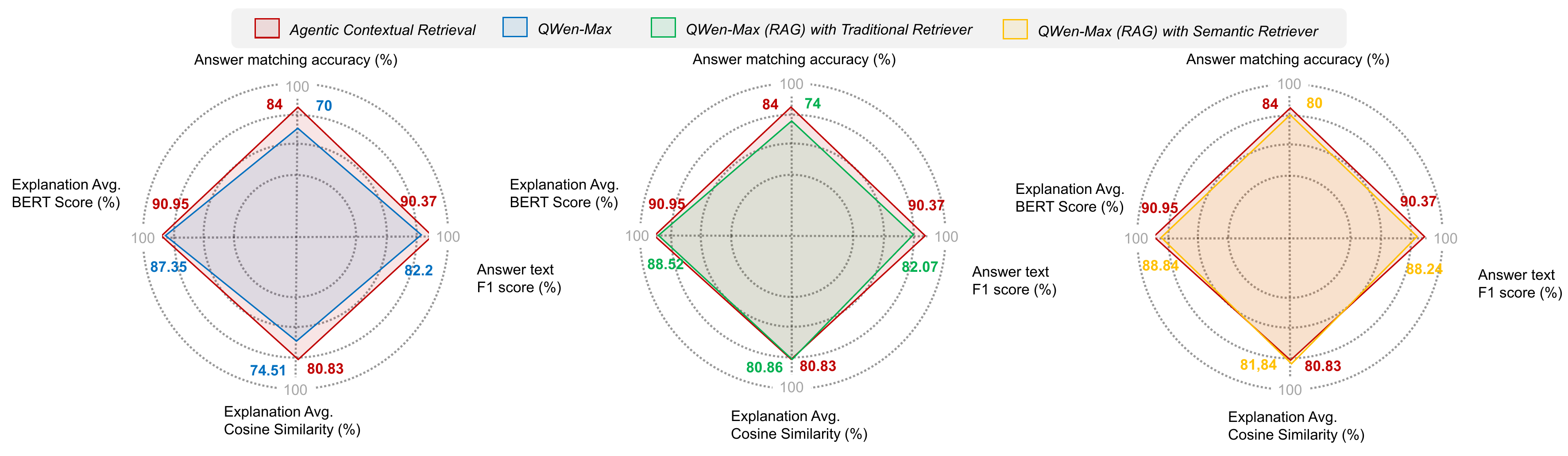}
\caption{Performance comparison of Agentic Contextual Retrieval against baseline methods, including QWen-Max without retriever, traditional retrieval, and semantic retrieval. }
\label{fig_PSNC_3}
\end{figure*}

\subsubsection{\textbf{Multi-Source Knowledge Retrieval}}
Following query optimization, the second step involves multi-source retrieval to ensure both completeness and relevance \textcolor{black}{in decision-making for network configuration and policy enforcement}.  Next, we integrate semantic vector-based retrieval with embedding models to extract key information from \textcolor{black}{3GPP specifications, network operation policies, and real-time telecom deployment scenarios.} Embedding models generate dense vector representations of text, enabling context-aware similarity search rather than relying on exact keyword matches \cite{shankar2024docetl}. To further improve accuracy, structured knowledge representations establish relationships between frequency bands, protocol parameters, and QoS metrics, refining query precision. Additionally, real-time retrieval from online repositories ensures access to the latest standardization updates. For instance, when retrieving information on "5G network slicing SLA guarantees," the system uses an embedding model to identify semantically relevant sections from TS 28.531 (Performance Assurance) and TS 28.554 (KPI Definitions) while incorporating recent case studies from network operators. 

\subsubsection{\textbf{Contextual Evidence Aggregation and Reasoning}} \textcolor{black}{Once relevant information is retrieved, the third step focuses on contextual evidence aggregation and reasoning, where multi-source knowledge is condensed into a structured and interpretable response. Given the vast amount of information available in telecom standardization, it is crucial to eliminate redundancy, enhance clarity, and ensure that the extracted content directly addresses the query \cite{li2025search}}. Specifically, we use an LLM-powered reasoning agent, which autonomously identifies the most relevant text segments in the retrieved content based on the reformulated query. The agent then synthesizes these segments into a concise, context-aware summary, ensuring that only the most important evidence is retained, and irrelevant or redundant information is discarded.  For example, in response to a question like ``What is the role of the serving network in fraud control?'', the retrieved information may contain detailed descriptions of charging functions, fraud detection, and policy enforcement. Instead of presenting all these details, the agent analyzes the content, extracts the core function of the serving network in fraud prevention, and generates a concise summary, emphasizing its role in real-time data collection and cost control. 

\subsubsection{\textbf{Decision-Making and Self-Validation}} 
\textcolor{black}{The final step involves a decision-making agent that simultaneously generates both the network action recommendations and justifications based on the optimized query and refined retrieval results.} This agent applies CoT reasoning to synthesize a structured response, ensuring that the explanation logically supports the answer by drawing from the retrieved evidence \cite{ayed2024hermes}. To enhance reliability, a self-reflection agent evaluates the generated response, critically reviewing both the answer and explanation for consistency, factual accuracy, and alignment with authoritative 3GPP standards. If inconsistencies, incomplete reasoning, or speculative conclusions are detected, the self-reflection agent challenges the response and triggers an iterative refinement loop. 

\subsection{Simulation}

\textbf{Simulation Settings:}
Our simulation is conducted using a structured retrieval and reasoning pipeline, integrating multiple knowledge sources and agent-driven query optimization. We employ Qwen2.5-Max\footnote{\url{https://huggingface.co/spaces/Qwen/Qwen2.5-Max-Demo}} as the base LLM, leveraging its advanced reasoning capabilities for telecom-related question-answering tasks. To evaluate retrieval performance, we selected 50 structured QA pairs related to 3GPP R18 from the TeleQnA dataset, which serves as the primary benchmark. For additional technical context, we use the 3GPP R18 dataset\footnote{\url{https://huggingface.co/datasets/netop/3GPP-R18}}. To ensure retrieval efficiency, we utilize FAISS\footnote{\url{https://github.com/facebookresearch/faiss}}, an indexing tool optimized for high-speed vector similarity search. The document processing workflow involves segmenting 3GPP standard documents into 1000-character chunks with a 100-character overlap, followed by embedding generation using Mpnet-base-V2\footnote{\url{https://huggingface.co/sentence-transformers/all-mpnet-base-v2}}, a transformer-based model trained for dense vector representations. To evaluate the effectiveness of the proposed Agentic contextual retrieval framework, we compare its performance against three baselines: (i) \textbf{Qwen-Max without Retriever}, representing a pure LLM-based approach, (ii) \textbf{Qwen-Max with Traditional Retriever}, utilizing standard retrieval-based augmentation, and (iii) \textbf{Qwen-Max with Semantic Retriever}, incorporating semantic embedding-based retrieval. The comparison is conducted across four key evaluation metrics, i.e., Answer Matching Accuracy, Answer Text F1 Score, Explanation BERT Score, and Explanation Cosine Similarity, as shown in Fig.~\ref{fig_PSNC_3}.

Fig.~\ref{fig_PSNC_3} demonstrates that Agentic contextual retrieval consistently outperforms all baseline methods across all evaluation metrics. In particular, the proposed framework achieves an answer matching accuracy of 84\% and an answer text F1 score of 90.37\%, surpassing the performance of semantic retrieval (i.e., 80\%) and traditional retrieval (i.e., 74\%), underscoring its effectiveness in generating precise and contextually relevant responses. This improvement is attributed to its dynamic multi-source retrieval, which integrates structured 3GPP standards with external knowledge repositories, query reformulation mechanisms, ensuring alignment with telecom-specific terminology, and a structured reasoning pipeline, which employs CoT decision-making and self-validation loops to enhance logical consistency and factual accuracy. Moreover, unlike conventional retrieval methods that rely on static document matching, Agentic contextual retrieval dynamically extracts, synthesizes, and validates multi-hop contextual information, significantly enhancing retrieval precision and response coherence. Furthermore, the explanation quality also benefits significantly from our approach, as evidenced by the Explanation BERT Score (i.e., 90.95\%) and Cosine Similarity (i.e., 80.83\%), both of which outperform alternative retrieval methods. These improvements stem from the framework’s ability to synthesize multi-source knowledge, apply structured reasoning, and iteratively refine responses through self-reflection mechanisms. In contrast, the semantic retrieval baseline, while effective at contextual retrieval, lacks robust reasoning capabilities and multi-turn validation, limiting its ability to handle complex telecom-specific queries.

\section{Future Directions}

\textbf{Security and Privacy in Retrieval-Augmented Networks:}
As agentic contextual retrieval frameworks increasingly rely on multi-source knowledge retrieval, ensuring data integrity, confidentiality, and adversarial robustness is critical. Future research should explore privacy-preserving retrieval techniques, such as federated retrieval, secure multi-party computation, and differential privacy-enhanced retrieval models, to mitigate risks associated with unauthorized data access and adversarial attacks in wireless and networking applications.

\textbf{Energy-Efficient and Low-Latency Retrieval Networking Architectures:}
Deploying LLM-driven agentic contextual retrieval frameworks in real-world wireless and networking environments requires optimized inference efficiency and low-latency retrieval mechanisms. Future studies could investigate mobile device-aware retrieval strategies, knowledge distillation for lightweight retrieval models, and edge-based retrieval deployment to minimize computational overhead while maintaining retrieval accuracy in resource-constrained environments, such as 5G edge nodes and IoT devices.

\textbf{Network-Aware Adaptive Retrieval for Real-Time Optimization:}
As telecom networks become increasingly complex and dynamic, retrieval systems must not only process knowledge efficiently but also adapt to real-time network conditions, congestion levels, and QoS constraints. Future research should explore network-aware retrieval architectures that dynamically adjust retrieval latency, query granularity, and resource allocation based on real-time network traffic and topology changes. Techniques such as reinforcement learning-based retrieval scheduling, adaptive caching, and traffic-aware retrieval pipelines could significantly enhance the responsiveness and efficiency in networking environments.

\section{Conclusion}
We have presented a forward-looking perspective on generative information retrieval-inspired intelligent communications and networking, emphasizing the role of retrieval in enhancing agentic AI for telecom systems. We have provided a comprehensive review of retrieval strategies. Additionally, we have reviewed recent retrieval-based studies in communications and networking. Then, we have introduced an LLM-based agentic contextual retrieval framework, which integrates multi-source knowledge retrieval, structured reasoning, and self-validation.

\bibliographystyle{IEEEtran}
\bibliography{mylib}

\end{CJK}
\end{document}